\documentclass[submission,copyright,creativecommons]{eptcs}
\providecommand{\event}{LFMTP 2021} 
\usepackage{underscore}           

\usepackage{enumitem}
\usepackage{multicol}

\usepackage{listings}
\usepackage[usenames]{xcolor}
\usepackage{Coq}
\usepackage{SML}

\DeclareMathAlphabet{\mathcal}{OMS}{cmsy}{m}{n}

\usepackage[colorinlistoftodos,prependcaption,textsize=tiny]{todonotes}

\title{SMLtoCoq: Automated Generation of Coq Specifications and Proof
Obligations from SML Programs with Contracts}

\author{
  Laila El-Beheiry \qquad Giselle Reis \qquad Ammar Karkour
  \institute{Carnegie Mellon University, Qatar}
  \email{loe@andrew.cmu.edu, giselle@cmu.edu, akarkour@andrew.cmu.edu}
}

\def\titlerunning{SMLtoCoq}
\def\authorrunning{L. El-Beheiry, G. Reis \& A. Karkour}

\begin{document}
\maketitle

\begin{abstract}
  Formally reasoning about functional programs is supposed to be
  straightforward and elegant, however, it is not typically done as a
  matter of course.
  Reasoning in a proof assistant requires ``reimplementing'' the code
  in those tools, which is far from trivial.
  SMLtoCoq provides an automatic translation of SML programs and
  function contracts into Coq.
  Programs are translated into Coq specifications, and function
  contracts into theorems, which can then be formally proved.
  Using the Equations plugin and other well established Coq libraries,
  SMLtoCoq is able to translate SML programs without side-effects
  containing partial functions, structures, functors, records, among
  others.
  Additionally, we provide a Coq version of many parts of SML's basis
  library, so that calls to these libraries are kept almost \emph{as
  is}.
  %
  %
\end{abstract}

\section{Introduction}

Programming language implementations are built for programming, so the
aim is to provide useful libraries and constructs to make writing
\emph{code} as easy as possible.
Proof assistants, on the other hand, are built for reasoning and as
such the aim is to make writing \emph{proofs} as easy as possible.
As much as functional programs look similar to (relational or
functional) specifications, if one wants to prove properties about an
implemented program, it is necessary to ``reimplement'' it in the
language of a proof assistant.
This requires familiarity with both the programming language and the
proof assistant, since not all programs are supported by reasoning
tools (e.g. non-terminating programs are typically forbidden), and not
all libraries are available in both tools.

In this work we present SMLtoCoq: a tool that \emph{automatically
translates SML code into Coq specifications}.
Moreover, we extend SML with \emph{function contracts which are
directly translated into Coq theorems}.
By using this tool, programs can be written using all conveniences of
a programming language, and their contracts (functions' pre and post
conditions) can be proved using the full power of a proof assistant.
Our target audience are programmers fluent in functional programming,
but with little expertise in proof assistants.
By having programs automatically translated into Coq code that looks
as similar as possible to SML, programmers can learn quickly how
program specifications look like, thus lowering the entry barrier for
using the proof assistant.

Even though both tools use a functional language, representing SML
programs in Coq is complicated due to their various differences.
The first immediate challenge is that SML programs may be partial and
have side-effects, while Coq programs/specifications need to be pure
and total. 
Another issue is that Coq requires recursive definitions to be
structurally decreasing so that termination is guaranteed. SML
programs can certainly diverge, and even if they terminate, it might
be because of a more complicated argument than a straightforward
structural induction.
Coq does have extensions -- such as the \coqe{Program} command -- that
are used to define non-structurally-decreasing, terminating recursive
functions. However, a proof of termination must be provided.
On top of these more fundamental problems, we have encountered many
small mismatches between the two languages, such as how records are
represented, and what type-checking can infer.

%
By using a number of Coq features and information available in SML's
abstract syntax tree (AST), SMLtoCoq is able to translate an extensive
fragment of \emph{pure} SML (i.e. without side effects),
including partial functions, records, and functors, among others.
The resulting Coq specification looks very similar to the original
code, so someone proving properties can easily map parts of
specifications back to their program counterpart.
Using the Equations library, we translate partial, mutually recursive,
and recursive functions out of the box. Non-terminating functions are
not accepted by Coq, but their translations type check.

Our contributions are:

\begin{itemize}

  \item We extend HaMLet's implementation of SML to parse and
  type-check function contracts, written as function annotations.
  These are included in the SML's AST and translated into theorems.
  
  \item We design and implement a tool, SMLtoCoq, that is able to
  translate pure SML programs and contracts into Coq specifications
  and theorems completely automatically.  Moreover, the translated
  code looks very similar to the original code.

  \item We implement in Coq the SML libraries\footnote{Some of these
  libraries are very pervasive and we needed their implementation to
  test the translation. This is why they were not translated using
  SMLtoCoq. See Section~\ref{sec:libs}.}: 
  \smle{INTEGER}, \smle{REAL}, \smle{STRING},
  \smle{CHAR}, \smle{Bool}, \smle{Option}, \smle{List},
  \smle{ListPair}, and \smle{IEEEReal}.

  \item We provide many examples of translated code, including a case
  study where we translate non-trivial SML code and prove properties
  on the Coq output. This aligns with the intended workflow for the
  tool: translate SML code, then prove properties in Coq.


\end{itemize}

SMLtoCoq can be found at:
\url{https://github.com/meta-logic/sml-to-coq/}

\section{Infrastructure}

In its core, SMLtoCoq implements a translation of SML's abstract
syntax tree (AST) into Gallina's (Coq's specification language) AST --
which is subsequently used to generate Gallina code. 
SML was chosen for being a language with an incredibly formal and
precise definition, which helped in understanding precisely what
fragments of the language were covered by our translation.
Coq was chosen for being an established and powerful proof assistant,
with many libraries and plugins available. In particular, it provides
the Equations library which was crucial for efficiently automating the
translation and generating correct code that looks close to
SML. 
%

SML's AST is obtained from HaMLet\footnote{\url{https://people.mpi-sws.org/~rossberg/hamlet/}}.
SML's implementation in HaMLet can be separated into three phases:
\emph{parsing}, \emph{elaboration}, and \emph{evaluation}. 
If a program's syntax is correct, \emph{parsing} will succeed
returning an AST with minimal annotation at each node, containing
only their position in the source code. 
Other well-formedness conditions (e.g. non-exhaustive or redundant
matches) and type-checking (i.e. static semantics) are computed during
\emph{elaboration}, which populates the annotations in the AST with
more information.  
\emph{Evaluation} will simply evaluate the program (i.e. dynamic
semantics).

We use the AST after the elaboration phase, when annotations contain
useful information such as inferred types and exhaustiveness of
matches. Such information is crucial for the generated Coq code. We call
evaluation to make sure the program executes correctly and terminates
(i.e. no exceptions raised or infinite loops entered)
before starting the translation, but the evaluation result is not used.
%
%
SMLtoCoq is implemented in SML.

\subsection{HaMLet}
\label{sec:hamlet}

HaMLet is an SML implementation whose goal is to be an
accurate implementation of the language definition~\cite{smldef} and
a platform for experimentation. We chose this implementation
for three main reasons:

\begin{enumerate}
  \item It is faithful to SML's definition, and all deviations and
  design choices are thoroughly documented.
  \item The resulting objects from each compilation step are easily
  accessible. Particularly, the AST after the elaboration phase could
  be obtained with a couple of function calls.
  \item Due to the detailed documentation, HaMLet could be easily
  modified as per the translation needs.
\end{enumerate}

HaMLet was extended with function contracts, written as
code annotations in the following syntax:

\begin{sml}
(!! f input ==> output;
    REQUIRES: exp1;
    ENSURES:  exp2;     !!)
\end{sml}

This contract must be placed immediately before \smle{f}'s declaration.
The first line includes the function name \smle{f} followed by the
variable bindings representing its \smle{input}.
These can be curried, uncurried, typed, or untyped, following the
syntax of function parameters in SML.
This is followed by \smle{output}, which is one named variable,
typed or not, representing the function's output.
The variable names in \smle{input} and \smle{output} are in the scope of
the contract only, and should not be confused with variables inside
the function.
\smle{REQUIRES} and \smle{ENSURES} are new keywords used for indicating
the pre and post condition of the function, respectively. 
They are followed by SML
boolean expressions \smle{exp1} and \smle{exp2}.
The type of these expressions is enforced by the type checker.
The variables used in \smle{input} and \smle{output} can be used in
\smle{exp1} and \smle{exp2}.
Variables in the function declaration are not available in the
contract, but \smle{exp1} and \smle{exp2} can use functions or
variables defined previously in the code.

In addition to changing HaMLet's lexer and parser, SML's AST was
modified to account for functions with contracts. 
The node representing function declaration was augmented to hold the
variable bindings and expressions from contracts.
If the function has no contracts, these fields are empty.

\medskip

Besides the implementation of function contracts, HaMLet needed to be
modified so that the translation would be as faithful as possible.
During parsing, constructs called \emph{derived forms} are
transformed into semantically equivalent code using other constructs.
This makes the core language smaller. For example, 
\smle{if-then-else} expressions are transformed into \smle{case}
expressions (which, in turn, are transformed into anonymous functions).
As a result, the AST computed after parsing would not have nodes for
\smle{if-then-else}, and its translation would result in a different
(although semantically equivalent) Coq specification.
We have changed HaMLet to skip the transformation of derived forms,
and added constructors to the AST to account for them.  The
constructors added were:

\begin{itemize}[noitemsep]
  \item Expressions%
    \vspace{-0.3cm}
    \begin{multicols}{2}
    \begin{itemize}[noitemsep]
      \item \smle{()} and \smle{(e_1, ..., e_n)}
      \item \smle{[e_1, ..., e_n]}
      \item \smle{case e of m}
      \item \smle{if e_1 then e_2 else e_3}
      \item \smle{e_1 andalso e_2} and \smle{e_1 orelse e_2}
      \item \smle{e_1 op e_2} (infix expressions)
    \end{itemize}
    \end{multicols}
    \vspace{-0.3cm}
  \item Patterns:
      \smle{()}, \smle{(p_1, ..., p_n)},
      \smle{[p_1, ..., p_n]}, and
      \smle{p_1 op p_2} (infix patterns)
  \item Tuple types: \smle{t_1 * ... * t_n}
  \item Function declarations (originally transformed into \smle{val rec}): \smle{fun id pats = e}
  \item Functor instantiation with inline specification (e.g.
        \smle{FuncID (structure <strdesc>)})
  \item Type definitions in signatures (e.g. \smle{type t = s},
  which was expanded to an inline signature)
\end{itemize}

SML's AST is annotated during elaboration phase. The annotation of the
new constructs can happen in one of two ways: (1) its equivalent form
is constructed, elaborated, and we use the resulting annotation
interpreted in the context of the derived form; or (2) a new
elaboration case is implemented for the construct. We have used a
combination of both approaches which incurred as few changes as
possible in the elaboration phase. The annotations produced by both
approaches are equivalent since they are added during the elaboration
phase which only considers the static semantics up to which the
derived form and its equivalent form are equivalent.

\subsection{Coq}

%


Coq's core language for writing specifications is called Gallina. To
be able to generate Gallina's AST, we have implemented it as a
datatype in our system. The implementation follows closely the grammar
of Gallina's
specification\footnote{\url{https://coq.inria.fr/distrib/current/refman/language/core/index.html}},
with a few small changes described in what follows. 

Extra term and pattern constructors were added for 
string, 
real, 
char,
tuple,
list,
unit,
and infix. 
SML expressions of those types are directly translated to Gallina
using these constructors, and they can be directly mapped to Coq
code either 
using common libraries or notations (e.g. \coqe{Datatypes}, \coqe{List}),
or Coq libraries that we have implemented to match SML libraries (e.g.
string, real, char).
New term constructors for product types, and boolean conjunction and
disjunction, were implemented for the same reasons as above. 
To be able to generate theorems, Gallina's AST includes the
\coqe{Prop} operators for conjunction, disjunction, equality, and
quantification.
Finally, \coqe{match} terms need to have a field indicating
exhaustiveness.

Some syntactical elements of Gallina were left out
since there is no corresponding SML code which generates them.
For example, 
type coercion (\coqe{:>}) and
``or'' patterns (\coqe{p_1 | p_2 => t}).

\paragraph{Equations}
Equations~\cite{eqtns} is a Coq plugin which allows a richer form of
pattern-matching and arbitrarily complex recursion schemes in the
setting of dependent type theory. 
Using this plugin, SMLtoCoq can generate concise and elegant code
that is visually similar to SML. The output is considerably improved
when compared to one generated in pure Gallina, and the derived proof
principles can be used to reason about the high-level code.
Among the main advantages of using Equations, is the ability to
define partial functions using dependent types. Domain restrictions
identified on the original SML function can be translated to a
dependent type, which is added as one of the function's arguments. The
resulting code has minimal overhead compared to its SML counterpart,
not needing extra default values or ad-hoc constructs to handle
unmatched cases.
In addition to that, Equations provides a simple interface to handle
non-trivial recursion where proofs of termination must be provided by
the user.

\subsection{Libraries}
\label{sec:libs}

SML's basis library is included in most of its implementations and
contains some of the more popular datatypes and functions. Since most
SML programs will make use of some part of the basis library, we have
implemented Coq equivalents to:
\smle{INTEGER},
\smle{REAL},
\smle{CHAR},
\smle{Bool},
\smle{STRING} (and partially \smle{StringCvt}),
\smle{Option},
\smle{List},
\smle{ListPair}, and
\smle{IEEEReal}.
They have the same interface as their SML counterpart, so function
calls to these libraries can be translated almost ``as is'' to
Coq. Most of the implementations build on existing Coq libraries, such
as \coqe{List}, \coqe{ZArith}, \coqe{Ascii}, \coqe{Bool},
\coqe{Floats}, \coqe{String} etc.

Several of SML's basis library functions raise \textbf{exceptions} for
a given set of inputs. However, Gallina is pure and does not have side
effects. In particular, it does not support exceptions. In order to
handle these cases, we return an \textbf{Axiom} instead of raising an
exception.
For example, the function \smle{List.hd} in SML's basis library raises
the exception \smle{Empty} if an empty list is passed to it. The
equivalent implementation of \smle{List.hd} in our libraries return
the axiom \coqe{EmptyException}, defined as:

\begin{coq}
Axiom  EmptyException : forall{a}, a.
\end{coq}

Note that this means it is not possible to \emph{catch} the exception,
and renders Coq inconsistent.
We are currently investigating possible solutions for exceptions, and
side-effects in general (see Section~\ref{sec:conc}).

A natural question to ask is why we have not used SMLtoCoq itself to
translate SML's libraries. The reason is purely of a practical matter:
function translation was being developed at the same time libraries
were being implemented in Coq, and one development informed the other.
Moreover, a lot of SML code relies on these libraries, so we needed
equivalent ones in Coq to test our translation.

%
%
%

\section{Translation} 
\label{sec:transl}

The translation from an SML's AST $\mathcal{S}$ into a Gallina's AST
$\mathcal{G}$ is defined inductively on $\mathcal{S}$. Depending on
the type of construct being translated, we rely on (local or global)
auxiliary contexts. 
We describe in this section the relevant parts of the translation,
including examples for each of them.
%
The code in all figures in this section is exactly the one used and
generated by SMLtoCoq, except for some line breaks and spaces removed
to fit the pages.
All the Coq code was tested with the following header, which imports
the libraries discussed in Section~\ref{sec:libs}, Equations, and sets
generalization of variables by default.

\begin{coq}
Require Import intSml.       Require Import listSml.
Require Import realSml.      Require Import stringSml.
Require Import charSml.      Require Import boolSml.
Require Import optionSml.    Require Import listPairSml.
Require Import notationsSml.

From Equations Require Import Equations.
Generalizable All Variables.
\end{coq}


More involved examples can be found in the \texttt{examples} folder in
the
repository\footnote{\url{https://github.com/meta-logic/sml-to-coq/tree/sml-to-coq-with-hamlet/examples}}.
In particular, we would like to highlight the \texttt{tree\_proof.v}
file which contains the translated code for functions on trees,
and proofs of non-trivial theorems about these functions.

\subsection{Overloaded operators}

Some comparison and arithmetic operators in SML are overloaded for
multiple types. For example, the equality check works for strings and
integers:
\smle{val b = "a" = "a" andalso 5 = 3}.

Boolean equality checks for string and integers in Coq, denoted by
\coqe{=?}, are defined in \coqe{string_scope} and \coqe{Z_scope},
respectively. However, the last opened scope shadows the first, and
trying to use both at the same time fails. For example:

\begin{coq}
Require Import ZArith. Open Scope Z_scope.
Require Import String. Open Scope string_scope.
Require Import Bool.

Fail Definition b := ("a" =? "a") && (5 =? 3).
\end{coq}

\noindent
fails because Coq expects 5 to be of type string.

Operation overload is solved using typeclasses. We have defined
several typeclasses for different sets of operators and instantiated
them with the types supported. 
We also defined notations for the operators to be the same as SML
whenever possible. 
For example, the typeclass below is instantiated for strings and
integers (among others):

\begin{coq}
Class eqInfixes A : Type := {
  eqb : A -> A -> bool;
  neq : A -> A -> bool
}.
Infix "="  := eqb (at level 70).
\end{coq}


\subsection{Records}
\label{sec:records}

A record is a set of named fields typed by record types, for example:

\begin{sml}
{name = "Bob", age = 42}: {name: string, age: int}
\end{sml}

In SML, records can occur as expressions, patterns, or types. When matching a
record pattern, the user can specify the names of relevant fields and
omit the remaining using ellipsis, as long as SML's type checker can
infer the full record type. For example:

\begin{sml}
fun getAge (r: {name: string, age: int }) = 
  case r of {age = x, ...} => x;
\end{sml}

Gallina supports record types,
however these must be declared using \coqe{Record}.  So the
\smle{getAge} function above could be:

\begin{coq}
Record rec := { name : string; age : Z }.
Definition getAge (r : rec) := match r with 
  {| name := _; age := x |} => x.
\end{coq}

There are three important points that must be taken into consideration
when translating records:

\begin{enumerate}
  \item Any record expression, type, or pattern in an SML declaration
  might require a \coqe{Record} declaration preceding the
  translation, and record types must be replaced by the \coqe{Record}
  identifier.

  \item The \coqe{Record} declaration automatically generates
  projection functions for each of the record's fields. As a result,
  field names cannot be reused in the code.

  \item There is no Gallina equivalent to SML's ellipsis when pattern
  matching records. So the translation must make all record fields
  explicit in patterns.
\end{enumerate}

To make sure all necessary records are declared in the translation, we
make use of a \emph{record context} associated with Gallina's AST.
This context is split into a local and a global part. The global
record context $\mathcal{R}_g$ contains all record types that already
have a declaration in the AST. The local record context
$\mathcal{R}_l$ is used in the translation of SML's declarations, and
starts empty. As the declaration is deconstructed, record expressions,
patterns, or types may be encountered. 
If there exists a record type in $\mathcal{R}_g$ or $\mathcal{R}_l$
such that the fields are the same, and their types are more general,
then the translation proceeds as usual.
If not, the type is stored in $\mathcal{R}_l$ using a fresh
name.
%
Once the declaration translation is done, the types from
$\mathcal{R}_l$ are translated into \coqe{Record}s occurring before
the declaration.  The content of $\mathcal{R}_l$ is then added to
$\mathcal{R}_g$.

To avoid name clashing, we modify the record fields' names by
prefixing it with the fresh name used for the record type. Each
time a record type, expression, or pattern is found, either its type
is found in the record context, or a new type is created. In both
situations we are able to tell what this prefix is, and rename the
fields accordingly.

Ellipsis on record patterns are resolved by looking into the
annotations after SML's elaboration phase. Since the record type must
be able to be inferred, this information can be extracted after
elaboration, and the pattern can be unfolded with all fields.

The result of translating a record type and a function on this type is
shown in Figure~\ref{fig:record}.

\begin{figure}[t]
\begin{center}
\begin{tabular}{c}
\begin{sml}
type r = { name : string, age : int }

fun isBob ({name = "Bob",...}: r) = true
  | isBob {...} = false
\end{sml} \\
\\
\hline
\hline
\\
\begin{coq}
Record rid_1 := { rid_1_name : string; rid_1_age : Z }.
Definition r := rid_1.

Equations isBob (x1: r): bool :=
  isBob {| rid_1_age := _; rid_1_name := "Bob" |} := true;
  isBob {| rid_1_age := _; rid_1_name := _ |} := false.
\end{coq}
\end{tabular}
\end{center}
\caption{Translation for records}
\label{fig:record}
\end{figure}

\subsection{Polymorphic Types}
\label{sec:polytypes}

The treatments of polymorphic values in SML and Coq are different. 
For example, in SML
\smle{val L = []} 
declares an empty list \smle{L} of type \smle{'a list}, where
\smle{'a} is a type variable.
This value can be safely used with instantiated lists: \smle{L = [3]}
is a well-typed boolean expression (which evaluates to \smle{false}).

In contrast, a ``polymorphic'' empty list can be declared in Coq in
(at least) two different ways:

\begin{coq}
Definition L1 := @nil.
Definition L2 {A : Type} := [] : list A.
\end{coq}

The types of the terms \coqe{L1} and \coqe{L2} \emph{as is} (i.e.
without annotations) are sligthly different:

\begin{coq}
L1 : forall A : Type, list A
L2 : list ?A where ?A : [ |- Type]
\end{coq}

The definition of \coqe{L1} looks more similar to what is written in
SML. However, if we want to use \coqe{L1} in other terms with
instantiated lists (such as \smle{L = [3]}), then we need to
write: \coqe{(L1 _) = [3]}\footnote{Here \coqe{\_} represents the type
variable \coqe{A} whose type is inferred by Coq.}.
To avoid adding the type parameter explicitly, we can use \coqe{L2},
which is implicitly interpreted as \coqe{L2 _} by Coq.
Indeed, the (type-)check \coqe{L2 = [3]} succeeds.

As a result, type variables are made explicit when translating
polymorphic SML value declarations so that these values can be used
\emph{as is} in the rest of the program, like the example of
\coqe{L2}.
This is done using a type variable context $\mathcal{T}$. 
The type variable context is always empty at the beginning of a
declaration's translation and, as the declaration is traversed,
``unknown'' types are added to the context. 
%
An unknown type becomes ``known'' when it is added to $\mathcal{T}$.
That is, when the translator encounters an expression \smle{e} with
unknown type $\alpha$, it adds $\alpha$ to $\mathcal{T}$. If a later
expression \smle{e'} has type $\alpha$, the translator treats this as
a known type, not changing the type variable context. At the end
of the translation, $\alpha$ is added as an implicit argument of the
resulting Coq definition, and the translation of the expression
\smle{e} is annotated with the type $\alpha$. 
For example, \smle{val L = []} is translated to:

\begin{coq}
Definition L {_'13405 : Type} := ([] : @list _'13405).
\end{coq}

\noindent
where \coqe{_'13405} is the name of the type variable determined by
HaMLet.

Note that this is only needed for value declarations. Functions on
polymorphic types do not need explicit type parameters since they can
be automatically generalized using \coqe{Generalizable All Variables}.

\subsection{Non-exhaustive Matches}

A very common practice when programming in SML is to use patterns for
values to deconstruct expressions. For example:
\smle{val x::l = [1,2,3]}
would result on value \smle{x} being bound to \smle{1} and \smle{l}
bound to \smle{[2,3]}. SML's interpreter will issue a \texttt{Warning:
binding not exhaustive}, but accepts the code. The warning makes sense
since it is usually not possible to tell before runtime if the expression on
the right will match the pattern (for example, when it is the result
of a function application). Non-exhaustiveness is indicated by a
flag in the declaration's annotation after elaboration.

Such declarations cannot be directly translated into Gallina because
\coqe{Definition}s cannot be patterns, only identifiers. Patterns are
accepted in \coqe{let ' pat := term} expressions, but \coqe{term} can
\emph{only} resolve to \coqe{pat} (i.e. its type has only one constructor).
The translation of non-exhaustive declarations is made exhaustive by
adding a default case resulting in a \coqe{patternFailure} axiom:
(the same strategy is used in~\cite{hs-to-coq}):
\coqe!Local Axiom patternFailure: forall {a}, a.!
We should note here that the default case will never be reached in the
translated code, as this would mean there was a \smle{Bind} exception
raised when evaluating the SML code.
As mentioned before, the SML code is evaluated before starting the
translation, and if it terminates abnormally (with an exception, for
example), SMLtoCoq terminates too.

In addition to that, top level declarations need to
be split into as many definitions as there are variables being bound.
This is done by recursively traversing the pattern and collecting the
variables. For each of them, a new Gallina definition is created. For
example, \smle{val x::l = [1,2,3]} becomes:

\begin{coq}
Definition x := match [1; 2; 3] with (x :: l) => x
                                   | _ => patternFailure
                end.
Definition l := match [1; 2; 3] with (x :: l) => l
                                   | _ => patternFailure
                end.
\end{coq}

Note that the structure of the match term is the same, apart from the
variable returned on the non-default case. If the declaration is
inside a \smle{let} block, it is translated into multiple nested
\coqe{let} blocks.

\subsection{Functions}
\label{sec:functions}

Unless a function is total and structurally decreasing at every
recursive call, it cannot be translated into \coqe{Fixpoint} (or
\coqe{Definition}, in case it is not recursive) directly. Most of the
problems we encountered in function translation could be solved using
the Equations plugin, which provides a powerful tool for defining
terminating functions via pattern-matching on dependent types.
Equations turned out to be more flexible and easier to use than Coq's
built-in \coqe{Program} command. 


\subsubsection{Pattern matching}
\label{sec:pat-matching}

Programming in SML typically makes extensive use of pattern-matching,
most common among which is pattern-matching on function inputs, for
example:

\begin{sml}
fun length [] = 0
  | length (x :: l) = 1 + length l
\end{sml}

Gallina, however, does not allow pattern-matching on function
parameters which means that the above function would -- in the best
case -- be translated to the following Coq code:

\begin{coq}
Fixpoint length {A : Type} ( id : list A ) :=
match id with
| [] => 0
| x :: l => 1 + length l
end.
\end{coq}

While this looks acceptable, the translation is complicated as the
number of (curried) parameters increases since \coqe{match} only deals
with one term at a time.
%
Equations allows the definition of functions by pattern matching on
the arguments without the need for an intermediary \coqe{match}
expression.
This enables SMLtoCoq to produce code that looks much more similar to the
corresponding SML code.  For example, the \smle{length} function
defined above translates to the following in Coq with Equations:

\begin{coq}
Equations length `(x1: @list _'14188): Z :=
  length [] := 0;
  length (x :: l) := (1 + (length l)).
\end{coq}

The main limitation associated with using Equations is that the
function's input and output types have to be explicit. 
This does not pose much of a threat since we have type information
from HaMLet's elaboration, and having them explicit does not affect
the semantics of the code.

\begin{figure}[t]
\begin{center}
\begin{tabular}{c}
\begin{sml}
(!! posAdd(x, y) ==> b;
    REQUIRES: x > 0 andalso y > 0;
    ENSURES: b > x andalso b > y;   !!)
fun posAdd(x, y) = x + y;
\end{sml} \\
\\
\hline
\hline
\\
\begin{coq}
Equations posAdd (x1: (Z * Z)
  posAdd (x, y) := (x + y).

Theorem posAdd_THM: forall x y b, posAdd(x, y)=b /\ ((x > 0) && (y > 0)) = true 
                                  -> ((b > x) && (b > y)) = true.
Admitted.
\end{coq}
\end{tabular}
\end{center}
\caption{Translation for function contracts}
\label{fig:contracts}
\end{figure}

\subsubsection{Contracts}

Deductive verification is a common way to do formal verification, 
which consists of generating mathematical proof obligations from the
code's specifications, then discharging these obligations using proof
assistants such as Coq or automated theorem provers.
Generating correct proof obligations is a crucial step in this process. 
To aid in this task, SMLtoCoq automatically translates function contracts (added
to HaMLet as explained in Section~\ref{sec:hamlet}) into Coq theorems.
A contract of the form:
\begin{sml}
(!! f input ==> output;
    REQUIRES: precond;
    ENSURES:  postcond;  !!)
\end{sml}

\noindent
is translated into:
\begin{coq}
Theorem f_Theorem: forall vars, (f input = output /\ precond = true) 
                                -> postcond = true.
\end{coq}

The theorem's \coqe{precond} and \coqe{postcond} are the translations of SML
boolean expressions \smle{precond} and \smle{postcond}, respectively.
As such, they have type \coqe{bool} and not \coqe{Prop}, hence the
need to use \coqe{= true} in the theorem statement. 
The quantified \coqe{vars} are the set of variables used in
\smle{input} and \smle{output}.
This theorem is placed after the function definition in the Gallina code
followed by \coqe{Admitted}, and the proof is left to the user.
An example of the resulting translation of a function with pre and
post conditions is shown in Figure~\ref{fig:contracts}.

\subsubsection{Mutual recursion}
\label{sec:mutual}

Figure~\ref{fig:mutual} shows the translation of a mutually recursive
type and function.
SML's \smle{and} construct maps nicely to Coq's \coqe{with}, which can
also be used for \coqe{Equations}.
One interesting thing to note about this example is how the
polymorphic types \coqe{evenList} and \coqe{oddList} need to be
annotated in Coq.
First of all, they take a type variable as an implicit
type due to the treatment of polymorphism explained in
Section~\ref{sec:polytypes}.
As a result, when this type is used and a type variable is passed
explicitly to it, it must be preceded by \coqe{@}.
Another thing noticeable in the translation is the presence of
\coqe{
Depending on the context, Coq cannot distinguish whether \coqe{*} is
the tuple type or \coqe{nat} multiplication, so the annotation
indicates to Coq that this is indeed a type.
%

\begin{figure}[t]
\begin{center}
\begin{tabular}{c}
\begin{sml}
datatype 'a evenList = ENil
                     | ECons of 'a * 'a oddList
and 'a oddList = OCons of 'a * 'a evenList

fun lengthE (ENil: 'a evenList): int = 0
  | lengthE (ECons (_, l)) = lengthO l
and lengthO (OCons (_, l)) = lengthE l
\end{sml} \\
\\
\hline
\hline
\\
\begin{coq}
Inductive evenList  {_a : Type} : Type := 
  | ENil  
  | ECons : (_a * @oddList _a)
with oddList  {_a : Type} : Type := 
  | OCons : (_a * @evenList _a)

Equations lengthE `(x1: @evenList _a): Z :=
  lengthE ENil := 0;
  lengthE (ECons (_, l)) := (lengthO l)
with lengthO `(x1: @oddList _a): Z :=
  lengthO (OCons (_, l)) := (lengthE l).
\end{coq}
\end{tabular}
\end{center}
\caption{Translation for mutually recursive type and function}
\label{fig:mutual}
\end{figure}

\subsubsection{Partial functions}
\label{sec:partial}

One of the powerful features of SMLtoCoq is its ability to handle
partial functions. 
While it is not possible to define partial functions in Coq as is,
restricting the translation to total code would be a big loss, not
only because partial functions are pervasive in programming, but also: 
(1) many times partial functions are defined with guarantees that the
function will not be called on non-valid arguments and 
(2) Coq's rich type system accepts preconditions on functions as part
of the function inputs.
To address this issue, we exploit Coq's powerful support for dependent
types to constrain the function domain. That is, we generate
preconditions for partial functions and add them as function inputs,
which Equations can then use to accept functions that handle the
subset of inputs that satisfies the generated preconditions.

A typical example is the {\tt head} function that returns the head of
the list \smle{fun hd (x::l) = x}, which is translated to:

\begin{coq}
Equations hd {A} (x1: list A) {H: \exists y1 y2, x1 = y1::y2}: A :=
  hd (x :: l) := x;
  hd _ := _. 
\end{coq}

\noindent
Note that the implicit parameter \coqe{H} ensures the function is only
called on non-empty lists, and is thus total.

In simple cases, Equations can automatically derive a contradiction
between the second case and the precondition and all obligations will
be discharged.
In more complicated cases, Equations would not be
able to derive the contradiction, and an obligation remains
for the user to solve\footnote{Coq accepts unsolved obligations at
  first, and a user can optionally ``admit obligations'' and resolve
  them when needed.}.

The preconditions can be generated by simply requiring that the input
parameters match one of the function's cases, for example:
\coqe{x = case1 \/ x = case2 \/ ... \/ x = casen}.
But this can lead to unnecessarily complicated preconditions in the
case of multiple input arguments and/or a function with multiple
branches. 
Instead, we use the AST's knowledge of exhaustive patterns to generate
preconditions that are considerably smaller than what would be
produced by this naive procedure. 
Our algorithm works by identifying \emph{generic} patterns, which
match any input.
For example, a single identifier or a constructor for a datatype with
one constructor would be generic patterns.
When preconditions are generated, generic patterns are eliminated
because they are not imposing restrictions on the input.
Consider this example:

\begin{sml}
fun hd_sum ((a,b)::l) ((a',b')::l') init = init + a + b + a' + b'
  | hd_sum ((a,b)::l) l'            init = init + a + b
  | hd_sum l          ((a',b')::l') init = init + a' + b'
\end{sml}

The naive search would produce the following proposition:

\begin{coq}
(\exists a b l, x1 = (a, b)::l /\ \exists a' b' l', x2 = (a', b')::l' /\ \exists init, x3 = init) \/
(\exists a b l, x1 = (a, b)::l /\ \exists l',       x2 = l'           /\ \exists init, x3 = init) \/
(\exists l,     x1 = l         /\ \exists a' b' l', x2 = (a', b')::l' /\ \exists init, x3 = init)
\end{coq}

Our procedure, however, produces:

\begin{coq}
(\exists y1 l,  x1 = y1 :: l /\ \exists y2 l', x2 = y2 :: l') \/
(\exists y1 l,  x1 = y1 :: l) \/
(\exists y2 l', x2 = y2 :: l')
\end{coq}

Note that this can be further simplified to 
\coqe{(\exists y1 l, x1 = y1 :: l) \/ (\exists y2 l', x2 = y2 :: l')}.
This requires the implementation of a simplification algorithm for
formulas, which we leave for future work.
The actual translation of the \smle{hd_sum} function is:

\begin{coq}
Equations hd_sum (x1: @list (Z * Z)
  {H: exists y1 y2, eq (x1) (y1 :: y2) /\ exists y1 y2, eq (x2) (y1 :: y2) \/ 
      exists y1 y2, eq (x1) (y1 :: y2) \/ exists y1 y2, eq (x2) (y1 :: y2)}: Z :=
  hd_sum ((a, b) :: l) ((a', b') :: l') := (((a + b) + a') + b');
  hd_sum ((a, b) :: l) l' := (a + b);
  hd_sum l ((a', b') :: l') := (a' + b');
  hd_sum _ _ := _.
\end{coq}

\begin{figure}[t]
\begin{tabular}{l||l}
\begin{sml}
signature PAIR =
sig
  type t1
  type t2
  type t = t1 * t2
  val default : unit -> t
end

structure IntString : PAIR =
struct
  type t1 = int
  type t2 = string
  type t = t1 * t2
  fun default () = (0, "")
end

functor Example (Pair : PAIR) =
struct
  val (a, b) = Pair.default ()
end

structure S = Example (IntString) 
\end{sml}
&
\begin{coq}
Module Type PAIR.
  Parameter t1 : Type.
  Parameter t2 : Type.
  Definition t := (t1 * t2)
  Parameter default : unit -> t.
End PAIR.

Module IntString <: PAIR.
  Definition t1 := Z.
  Definition t2 := string.
  Definition t := (t1 * t2)
  Equations default (x1: unit
    (Z * string)
    default tt := (0, "").
End IntString.

Module Example ( Pair : PAIR ).
  Definition a := 
    match (Pair.default tt) with
    (a, b) => a end.
  Definition b := 
    match (Pair.default tt) with
    (a, b) => b end.
End Example.

Module S := !Example IntString.
\end{coq}
\end{tabular}
\caption{Translation for the module system}
\label{fig:module}
\end{figure}

\subsection{Structures, Signatures, and Functors}

An SML program can be divided into structures -- with each structure
comprising a collection of components, (i.e. datatypes, types, and
values). 
A structure can ascribe to a signature, which acts like an interface.
In addition to structures and signatures, SML also provides
\emph{functors}, which are essentially structures parametrizes by
other structures~\cite{prog-in-sml}.
Gallina has also a module system that provides similar mechanisms for
structuring programs. While Gallina's modules and module types
conveniently match SML's module language, there is one major
syntactical limitation: inline structures and signatures.

Structure and signature expressions in SML can be either top-level or
inline. Top-level structure and signature expressions in SML have a
single format, where \smle{S} is the component's name:

\begin{sml}
structure S  = structure_exp
signature S  = signature_exp
\end{sml}

Inline structure expressions occur as functors'
parameters:

\begin{sml}
structure S = functor F (structure_exp)
\end{sml}
\noindent
and inline signature expressions can occur in any of the
following ways:

\begin{sml}
structure_exp : signature_exp
structure_exp :> signature_exp
include signature_exp
structure S : signature_exp
functor F (structure S : signature_exp) = structure_exp
\end{sml}
\noindent
Note that inline structures and signatures are unnamed.


In Gallina, inline modules and module types are not allowed; they must
be replaced by identifiers. For example, the following attempt of
instantiating a \coqe{ListOrdered} module
directly in the parameter of the \coqe{Dict} module:
\coqe{Module D := Dict(ListOrdered(IntOrdered))}
fails.
%
Therefore, we distinguish between the translation of top-level
expressions and inline expressions.
The translation of inline expressions uses a structure/signature
context $\Sigma$. This context is initially empty and gets populated
with ASTs for inline signature and structure declarations as they are
discovered in (possibly nested) inline expressions. Each declaration
in this context is assigned a new name, and this name is used for
instantiating the Coq module.
An example of how modules are translated is shown in
Figure~\ref{fig:module}.

\subsection{Infix Functions}

SML allows the declaration of infix functions via the \smle{infix} and
\smle{fun op} constructs.
HaMLet keeps track of infix functions in the \emph{infix environment},
which is returned after parsing, together with the
AST.
As a result, the \smle{infix} declaration is not part of the AST.
When an infix function is used as a prefix, it is preceded by
\smle{op}, and this information is available in the AST.
Infix functions can be declared in Coq using \coqe{Notation} or
\coqe{Infix}.
For our translation we need to use \coqe{Notation} because
\coqe{Infix} assumes the function to be curried, while in SML infix
functions are always uncurried.

Once a function declaration of the shape \smle{fun op f} is found in
SML's AST, SMLtoCoq checks if \smle{f} in the infix environment, in
which case it was declared as an infix function\footnote{Note that SML
allows non-infix functions to be declared using \smle{fun op}.}.
If \smle{f} is infix, it creates two additional Gallina sentences to
be placed after the function definition. 
First, it defines \coqe{opf} as \coqe{f} to be used when \smle{op f}
is used in the SML code.
Secondly, it defines the \coqe{Notation "x 'f' y"} so that \coqe{f}
can be used in infix form from this point onward.
See an example in Figure~\ref{fig:infix}.
Similar to records, SMLtoCoq uses a context to keep track of which
functions have infix notations in the Gallina AST.

\begin{figure}
\begin{center}
\begin{tabular}{l||l}
\begin{sml}
infix F
fun op F (x, y) = x*x + y
val f = op F
val x = 5 F 2
val y = op F (2, 3)
\end{sml}
&
\begin{coq}
Equations F (x1: (Z * Z)
  F (x, y) := ((x * x) + y).
Definition opF := F.
Notation "x 'F' y" := (F (x, y)) 
         (left associativity, at level 29).
Definition x := (5 F 2).
Definition y := (opF (2, 3)).
\end{coq}
\end{tabular}
\end{center}
\caption{Translation for infix functions}
\label{fig:infix}
\end{figure}

%

\subsection{Typed patterns}

SML supports types in patterns at any level. The same does not hold
for Gallina. For example:

\begin{coq}
Definition f := fun '((x, y) : nat * nat) => x + y.
Fail Definition f := fun '((x: nat, y: nat)) => x + y.
Fail Definition f x := match x with | x : nat => 1 | _ => 0 end.
\end{coq}

Typing the pattern in the first definition is accepted, however,
types ``inside'' the pattern as in the second definition are rejected.
Also, typing patterns in \coqe{match} expressions is not accepted at
all. 

If we want to retain the same explicit types as in the SML code, it is
possible to extract types nested in patterns to the top level. 
However, this may include other types that were not explicit.
Also, we would need to identify precisely when top-level patterns are
not supported, and find an alternative to make the types explicit. 
For \coqe{match} expressions, for example, one can type the expression
being matched as opposed to the pattern.
Due to these conflicts, our design choice was to ignore types in
patterns. 
Most of the times this is not problematic as Coq is able to infer the
correct types. 
Moreover, function types are already explicit since \coqe{Equations}
requires it.

%


\section{Related Work}

Closest to our approach are hs-to-coq~\cite{hs-to-coq}, which
translates Haskell code into Coq specifications, and
coq-of-ocaml\footnote{\url{https://clarus.github.io/coq-of-ocaml/}},
which translates OCaml code into Coq specifications.
%
%
Even if these projects look (superficially) the same, an important
difference is that our main goal is to lower the entry barrier into
interactive theorem proving (in particular, Coq) for those that are
familiar with functional programming (in particular, in SML). 
As such, we aim for a translation that looks as similar as possible to
the SML code, which is obtained using the Equations plugin. We note
that Equations is not used in hs-to-coq or coq-of-ocaml, and getting
similarly looking code does not seem to be a priority in those
projects, which are more focused on the formal verification of large
codebases. 
Another distinguishing feature is the implementation of contracts for
functions, which is not available in Haskell or OCaml. 

%

As mentioned, SMLtoCoq uses the Equations plugin, while both hs-to-coq
and coq-of-ocaml translate functions to \coqe{Definition}s or
\coqe{Fixpoint}s.
As such, they look quite different (and less elegant) than their
Haskell or OCaml counterparts.
For example, definitions of mutually recursive functions in hs-to-coq
require the bodies of the functions to be repeated for each
definition, a problem that is solved using mutually defined
\coqe{Equations} (Section~\ref{sec:mutual}).
We find that the Equations plugin helps in obtaining more
aesthetically pleasing functions.
Haskell functions are defined via cases, like in SML, but this is not
supported in Gallina (as explained in Section~\ref{sec:pat-matching}).
Therefore, hs-to-coq must create intermediate names for the arguments
to be used in \coqe{match} expressions. SMLtoCoq avoids this, again
via the Equations plugin.
When it comes to partial functions, both hs-to-coq and coq-of-ocaml
add a default case on \coqe{match} expressions, and use an
\coqe{Axiom} as its return value.  
In contrast, SMLtoCoq generates the domain restriction as a dependent
type to be used in the \coqe{Equations} and avoids the need for a (yet
another) \coqe{Axiom} (see Section~\ref{sec:partial}).
We find this is a particularly elegant solution, as it reduces the
amount of inconsistent axioms that need to be used.
%

Other smaller differences include the treatment of built-in types and
records. 
While hs-to-coq translate types such as \texttt{Int} into their own
implementation \coqe{GHC.Types.Int}, SMLtoCoq tries to leverage Coq's
types as much as possible, translating \smle{int} into Coq's \coqe{Z}.
Records in hs-to-coq are translated into \coqe{Inductive} types with
associated projection functions. SMLtoCoq uses Coq's built-in
\coqe{Records}, having no need to declare projection functions
explicitly (see Section~\ref{sec:records}).
The treatment of these constructs in coq-of-ocaml is similar to ours.
It is worth noting that coq-of-ocaml curries all constructors, while
SMLtoCoq and hs-to-coq retain the user defined datatypes as
faithful as possible to the original definition.

%



CFML~\cite{cfml} is a tool for verifying Caml programs based on the
so-called \emph{characteristic formulae}. These formulas are derived
automatically from the program (without the need for annotations), and
describe its behaviour. The resulting formula can be proved using Coq. 
CFML's newest version uses separation logic to reason about OCaml
code~\cite{cfml2}. 
Differently from SMLtoCoq, CFML ``translates'' functions into
specification lemmas in Coq, in the style of Hoare triples. 

%


The Why tool~\cite{why3} encompasses WhyML, and ML-like language with
support for annotations, and the verifier Why3. Why3 leverages several
automated and interactive theorem provers to discharge proof
obligations coming from WhyML code as automatically as possible.
Even if the language resembles our use of contracts in SML, the goal
is not the same. SMLtoCoq uses solely Coq for verification, and does
not aim at automation.

\smallskip

SMLtoCoq translates SML programs into Coq specifications for
reasoning. Going in the other direction, Coq has an extraction
mechanism which exports specifications to OCaml, Haskell, or Scheme.
Similarly, F$^*$~\cite{fstar} is an OCaml-like functional language
which allows the programmer to state and prove lemmas about the code.
After verification, programs can be extracted in OCaml, F\#, C, WASM,
or ASM.
To use those tools the programmer needs to have expertise with proof
assistants in advance. SMLtoCoq assumes a greater fluency in
programming and less in theorem proving, thus enabling the user to
write programs in their comfort zone, and later experiment with
proving properties in a proof assistant. We believe this direction
helps beginners in Coq understanding how program specifications
would look like.


\section{Conclusion \& Future Work}
\label{sec:conc}

We have described SMLtoCoq, a tool for automatically generating Coq
specifications from SML programs. To the best of our knowledge, this
is the first tool of its kind. SMLtoCoq is able to handle a
considerable fragment of SML, including constructs that are not
trivially translated into Gallina, such as partial functions,
structures, and records.
Additionally, we have implemented contracts for functions and their
translation into Coq theorems.
We have also ported a big part of SML's basis library into Coq, so
that the code can be translated with the minimum amount of
modifications.
Using the resulting translation the user is able to prove properties
about their code using Coq.

We plan to improve and extend SMLtoCoq in several ways.

\vspace{-0.5cm}
\begin{paragraph}{Simplification of automatically generated pre-conditions}
As mentioned in Section~\ref{sec:partial}, the automatically generated
preconditions for functions can be further simplified by using logical
equivalences.
Since the result is always a proposition in disjunctive normal
form, we plan to improve our procedure by converting that to
conjunctive normal form and applying SAT heuristics for
simplifying propositions. 
\end{paragraph}

\vspace{-0.5cm}
\begin{paragraph}{Functions inside let blocks}
One of the drawbacks of using the Equations library is that
\coqe{Equations} is a top-level declaration. In SML, functions can be
declared nested inside let blocks, for example:
\begin{sml}
fun f x = let fun g y = y + 1 in g x end
val n = let fun h x = x + 1 in h 7 end
\end{sml}

The definitions of \smle{f} and \smle{g} could be declared mutually
using Equations' \coqe{where} construct. However, if the nesting depth
is bigger than 2, the translation would flatten the structure, which
could turn out to be problematic if some functions have the same name.
The definition of \smle{h}, on the other hand, cannot be translated
into \coqe{Equations}. In this case, we need to define a new
translation using Coq's native \coqe{let (fix)}.
\end{paragraph}

\vspace{-0.5cm}
\begin{paragraph}{Non-trivial recursion}
One of the fundamental differences between SML and Coq is that Coq
only accepts terminating functions. Among those, functions that are
non-trivially terminating must be accompanied by a proof of
termination to be accepted. At the moment, SMLtoCoq translates all
functions correctly, but those that need termination proofs cannot be
compiled in Coq.

For example, the \coqe{div_two} function:
\begin{coq}
Equations div_two (n : nat) : nat :=
div_two 0 := 0;
div_two 1 := 0;
div_two n := 1 + div_two (n / 2) .
\end{coq}
is terminating, but not trivially since the recursive call is not on
the predecessor of \coqe{n}, but on \coqe{n / 2}.
To make Coq accept this function, we can annotate it with the
well-founded inductive measure to use for the recursive calls. In this
case it is the simple \coqe{lt} function:
\begin{coq}
Equations div_two (n : nat) : nat by wf n lt :=
div_two 0 := 0;
div_two 1 := 0;
div_two n := 1 + div_two (n / 2) .
\end{coq}
This generates the proof obligation $n / 2 < n$ for $n \geq 2$ that
needs to be proved by the user.

Using the \coqe{by} annotation, we can also have Coq accept
non-terminating functions, but at the cost of an inconsistent axiom
and admitted obligations.
\begin{coq}
Local Axiom indMeasure: forall {a}, a -> nat.

Equations loop (x1: Z): Z by wf (indMeasure x1) _ :=
  loop x := (loop1 ((x + 1))).
Admit Obligations.
\end{coq}

We are investigating ways to determine termination information for SML
functions, ideally analogous to Coq's termination check.
It is unlikely we can automatically figure out the correct inductive
measure to use in the annotation (if there is any), but we can at
least have functions that compile, and leave it up to the user to
remove the use of inconsistent axioms and prove the obligations when
possible.
\end{paragraph}

\vspace{-0.5cm}
\begin{paragraph}{Side-effects}
Two language features that we have largely (and reasonably) ignored
were exceptions and reference cells. These operations involve side
effects and, naturally, have no easy correspondence in Coq.
Fortunately, dealing with side-effects in pure functional settings is
a well-studied problem~\cite{ynot,interaction-trees,free-spec} and
there are different solutions, including Coq libraries, that we could
adapt to translate effectful SML code.
\end{paragraph}

\vspace{-0.5cm}
\begin{paragraph}{Correctness}
We would like to formally prove that our translation for SML into
Gallina is correct, which would ultimately guarantee that the
reasoning about the code in Coq translates to its SML source. To that
end, we started to formalize our translation as a derivation system. 
Our goal is to show a simulation theorem between the SML source and
its translation, using both languages' evaluation semantics. We will
start with a small, purely functional, core of both languages, and
extend from there.
\end{paragraph}


\bibliographystyle{eptcs}
\bibliography{references}


\end{document}